# Analysis of nonclasscial features in a coupled macroscopic binary system

Byoung S. Ham
Center for Photon Information Processing, School of Electrical Engineering and Computer Science,
Gwangju Institute of Science and Technology
123 Chumdangwagi-ro, Buk-gu, Gwangju 61005, S. Korea
(Submitted on July 28, 2020; bham@gist.ac.kr)

Nonclassical phenomena of quantum mechanics such as anticorrelation and photonic de Broglie waves (PBWs) have been recently understood as a special case of coherence optics with a particular phase relation between orthogonal bases composing a classical system. Such a macroscopic understanding of nonclassical features has also been confirmed experimentally for a coherence version of PBWs in a doubly-coupled Mach-Zehnder interferometer (MZI). Here, a multi-coupled MZI system is analyzed and discussed to obtain a general understanding of the nonclassical feature using tensor products of binary bases of a classical system. This analysis should intrigue a fundamental question of quantumness or nonclassicality limited to a microscopic world of a single photon or a single particle.

## 1. Introduction

Quantum superposition is the bedrock of not only coherence optics of interference and diffraction [1], but also quantum mechanics of nonclassical features such as anticorrelation [2-10], Franson-type nonlocal correlation [11-15], photonic de Broglie waves (PBS) [16-20], and entangled photon/spin pairs [21,22]. A typical example of superposition is a Young's double slit or a Mach-Zenhder interferometer (MZI), where such coherence optics has been applied for both macroscopic (coherent lights) [1] and microscopic (single particles) regimes [11-15,23]. The key concept of quantum superposition is indistinguishability between bipartite entities in a coherence regime [24] as shown for a single photon itself via self-interference in a microscopic world [23]. For this indistinguishability, phase coherence plays a key role for quantum superposition, resulting in the uncertainty principle. Unlike the energy-time bin uncertainty relation in the microscopic world as a particle nature of quantum mechanics [2-21], a phase relation has been recently understood as the key concept of quatumness or nonclassicality [25], where such quantumness can also be induced by macroscopic coherence optics [26-29]. Here, the phase relation between bipartite entities implies the origin of the nonclssical features, where the bipartite-based quantumness can be extended into a macroscopic world such as for Schrodinger's cat [30-35].

Recently, a brave interpretation has been applied for the Hong-Ou-Mandel (HOM) effect of photon bunching or anticorrelation to understand the fundamental physics of quantumness [25]. Unlike conventional understanding of the particle nature in quantum mechanics based on the energy-time bin relation [2-22], a simple theory of coherence optics has been successful at inducing the same results in a classical system, where a relative phase between the bipartite is the key concept for the quantum feature of anticorrelation [25-29]. In this theory, the so-called HOM dip is supposed to be sinusoidally modulated as a function of the phase difference between two input photons within coherence time, where the conventional post-measurements are rather a minor process due to much delicate or sensitive phase effect on coherence as presented in Franson-type nonlocal correlation [11]. The Franson-type nonlocal correlation between remote bipartite entities has also been understood in terms of coupled coherence physics between noninteracting MZIs via coincidence detection [27]. Finally, a coherence version of PBW [28] has been proposed recently under the name of coherence de Broglie waves (CBW) in an asynchronously coupled double Mach-Zehnder interferometer (ACD-MZI) system to demonstrate coherence optics-based quantum features applicable to quantum sensing and quantum metrology [36-43].

Very recently, CBW has been experimentally demonstrated in a pure classical regime of the ACD-MZI [29], where MZI has been implemented as a quantum device as discussed already for anticorrelation [25] and Franson-type nonlocal correlation [27]. Here, a general analysis of CBW is presented to give a better understanding of how such a nonclassical feature can be created in a pure classical system of MZIs. For this, quantum superposition between two paths of an MZI is investigated in a coupled system, where higher order phase-basis creation plays a major role. Here, we demonstrate that higher order phase bases are the origin of quantum features in CBW.

Furthermore, a cavity CBW [37] is also investigated for potential applications of coherence quantum sensing such as inertial navigation [38,39] and quantum lithography [40-43]. Here, the cavity CBW is also compared with an optical cavity to present their difference in fundamental physics.

## 2. Results

*Analytical approach*

Figure 1 shows a schematic of the coupled ACD-MZI for CBWs in ref. 28. Here, ACD is an asymmetrical phase relation between consecutive MZIs, i.e., $\varphi_{CD} = -\varphi_{AB}$, where $\varphi_{ij} = \varphi_j - \varphi_i$ (see the blue and green dotted boxes). Here, the sign "$-$" stands for a π–phase shift to the given phase $\varphi_{ij}$ in harmonic oscillation of light fields. For the phase basis of a single MZI, we redefine it from $\{0, \pi\}$ to $\{0, \varphi\}$ for the ACD-MZI. The multiply coupled MZI system in Fig. 1 is analyzed with an n-ordered tensor matrix notation of the phase bases for $\psi = 0$, where the π–basis of ψ induces a π–phase shift in the path superposition [28]:

$$\begin{bmatrix} 0 \\ \varphi \\ \vdots \\ \varphi^q \end{bmatrix}_{qx1} = \begin{bmatrix} 0 \\ \varphi \end{bmatrix}_1 \otimes \begin{bmatrix} 0 \\ \varphi \end{bmatrix}_2 \cdots \otimes \begin{bmatrix} 0 \\ \varphi \end{bmatrix}_q . \qquad (1)$$

In equation (1), q stands for the number of MZIs. Thus, the output of the q-coupled MZI system of Fig. 1 has $2^q$ combinations with $(1 + q)$ phase bases in terms of $\varphi^j$ $(j = 0,1,2, \ldots, q)$. Because the phase φ applies to the exponent of harmonic waves, the output field's phase can be represented by $e^{iq\varphi_j}$ (j=0,1,2…). In other words, the output phase basis varies along with q, where each φ basis is uniformly synchronized for all MZIs. Thus, the phase of the $\varphi^q$ basis in equation (1) is represented by $\varphi^q = (2\pi/2q)m$, where m=0,1,2...

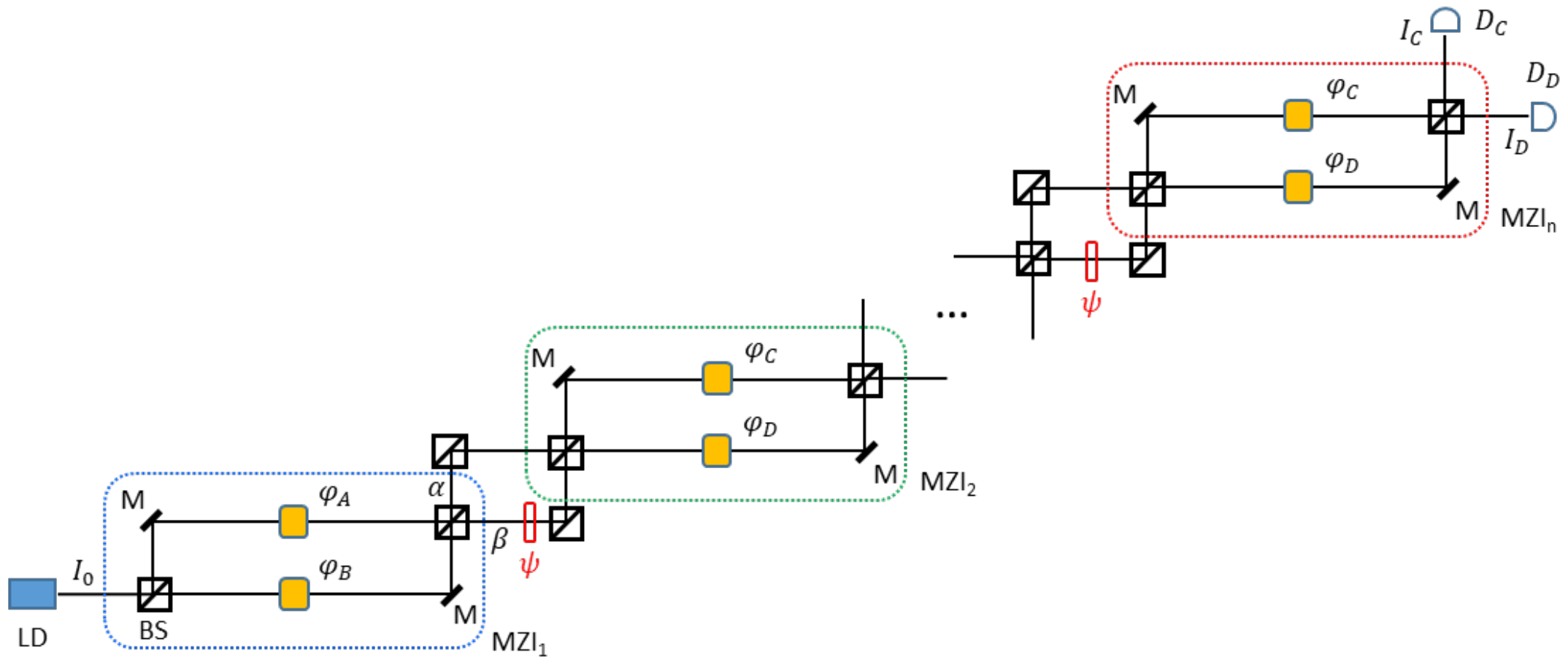

**Figure 1.** An n-folded MZI system. Each $\varphi_j$ has two bases, $\varphi_j \in \{0, \pi\}$. The phase ψ satisfies $\psi \in \{0, \pi\}$. LD, laser diode; I, isolator; C, optical cavity; M, mirror; BS, beam splitter; $\varphi_j$, phase in each path; induced by path-length or frequency difference; ψ, control phase; D, detector; $I_0$, input field intensity.

To understand the coupled MZI system of Fig. 1, equation (1) is investigated with ordered bases of $\varphi^q$ as follows:

(i) For q=2 in a doubly coupled MZI

$$\begin{bmatrix}0\\ \varphi\end{bmatrix}_1 \otimes \begin{bmatrix}0\\ \varphi\end{bmatrix}_2 = \begin{bmatrix}0\\ \varphi\\ \varphi\\ \varphi^2\end{bmatrix}. \tag{2}$$

In a doubly coupled MZI system, the phase of the q^th^ ordered basis is simply qφ because each phase basis is superposed in the output field as $(e^{i\varphi})^q$. Thus, equation (2) implies CBW [28] whose eigenstates are $\varphi^2$, φ, and 0.

We now define the ACD-MZI scheme as n=1 (q=2) in Fig. 1 (see both blue and green boxes). The basic building block of ACD-MZI is analyzed using matrix representations as follows. For this, the asymmetric coupling is set at $\varphi_{CD} = -\varphi_{AB} \equiv -\varphi$ for $\psi \in \{0, \pi\}$ (for a symmetric case, see Section of 1 of the Supplementary Information):

$$\begin{bmatrix}E_A\\ E_B\end{bmatrix} = [-M][\Psi][M]\begin{bmatrix}E_0\\ 0\end{bmatrix},$$

$$\frac{1}{4}\begin{bmatrix}(1-e^{-i\varphi})(1-e^{i\varphi}) - e^{i\psi}(1+e^{-i\varphi})(1+e^{i\varphi}) & i\left[(1-e^{-i\varphi})(1+e^{i\varphi}) - e^{i\psi}(1-e^{i\varphi})(1+e^{-i\varphi})\right]\\ i\left[(1-e^{i\varphi})(1+e^{-i\varphi}) - e^{i\psi}(1-e^{-i\varphi})(1+e^{i\varphi})\right] & -(1+e^{-i\varphi})(1+e^{i\varphi}) + e^{i\psi}(1-e^{-i\varphi})(1-e^{i\varphi})\end{bmatrix}, \tag{3}$$

where $[-M] = [BS][-\varphi][BS]$, $[M] = [BS][\varphi][BS]$, $[BS] = \frac{1}{\sqrt{2}}\begin{bmatrix}1 & i\\ i & 1\end{bmatrix}$, $[\varphi] = \begin{bmatrix}1 & 0\\ 0 & e^{i\varphi}\end{bmatrix}$, $[-\varphi] = \begin{bmatrix}1 & 0\\ 0 & e^{-i\varphi}\end{bmatrix}$, and $[\Psi] = \begin{bmatrix}1 & 0\\ 0 & e^{i\psi}\end{bmatrix}$. To solve equation (3), two different values of the ψ −bases are separately considered:

(a) For $\psi = 0$,

$$\begin{bmatrix}E_A\\ E_B\end{bmatrix} = -\frac{1}{2}\begin{bmatrix}e^{i\varphi} + e^{-i\varphi} & -i(e^{i\varphi} - e^{-i\varphi})\\ i(e^{i\varphi} - e^{-i\varphi}) & e^{i\varphi} + e^{-i\varphi}\end{bmatrix}. \tag{4}$$

Equation (4) is rewritten as:

$$\begin{bmatrix}E_A\\ E_B\end{bmatrix} = (-1)\begin{bmatrix}\cos(\varphi) & \sin(\varphi)\\ -\sin(\varphi) & \cos(\varphi)\end{bmatrix}\begin{bmatrix}E_0\\ 0\end{bmatrix}. \tag{5}$$

The corresponding intensities are as follows, respectively:

$$I_A = \frac{1}{2}[1 + \cos(2\varphi)], \tag{6}$$

$$I_B = \frac{1}{2}[1 - \cos(2\varphi)], \tag{7}$$

where $I_j = E_j E_j^*$. Equations (6) and (7) show typical CBW in ACD-MZI for n=1 [28,29]. For $n \geq 1$, the following relation is obtained straightforwardly from equation (5) by matrix multiplications:

$$\begin{bmatrix}E_C\\ E_D\end{bmatrix}^n = (-1)^n\begin{bmatrix}\cos(n\varphi) & \sin(n\varphi)\\ -\sin(n\varphi) & \cos(n\varphi)\end{bmatrix}\begin{bmatrix}E_0\\ 0\end{bmatrix}. \tag{8}$$

The corresponding intensities are as follows, respectively:

$$(I_C)^n = I_0\cos^2 n\varphi, \tag{9}$$

$$(I_D)^n = I_0 sin^2 n\varphi. \tag{10}$$

(b) For $\psi = \pi$,

$$\begin{bmatrix}E_C\\ E_D\end{bmatrix} = \begin{bmatrix}1 & 0\\ 0 & 1\end{bmatrix}\begin{bmatrix}E_0\\ 0\end{bmatrix}. \tag{11}$$

Equation (11) is for the case of unconditionally secured classical key distribution (USCKD) as an identity relation between the input and output [26]. This identity relation is the same as the $\varphi$ −symmetric case with $\psi = 0$ (see Section 1 of the Supplementary Information). For the tensor notation in equation (2), equation (11) is represented as:

$$\begin{bmatrix} 0 \\ \varphi \end{bmatrix}_1 \otimes \begin{bmatrix} 0 \\ -\varphi \end{bmatrix}_2 = \begin{bmatrix} 0 \\ -\varphi \\ \varphi \\ 0 \end{bmatrix}, \tag{12}$$

where the " − " sign of $\varphi$ in equation (12) is due to the $\pi$ −phase shift by the control phase $\psi$. Like equation (11), equation (12) also shows the identity relation with the same phase bases. Here, the sign in equations (5) and (12) does not alter the intensity. Although two different cases of (a) and (b) result in distinct outputs, they root in the same physics of nonclassical CBW, where the identity relation is an extreme case of CBW without modulation (see Section 1 of the Supplementary Information for the $\psi$ −dependent amplitudes) [29]; this will be discussed below in Fig. 2.

To investigate ACD-MZI in Fig. 1 for the ordered phase bases, equation (2) needs to be redefined as:

$$\begin{bmatrix} 0 \\ \varphi \\ \varphi \\ \varphi^2 \end{bmatrix} \to (1,2,1), \tag{13}$$

where the parentheses represent coefficients of the ordered phase basis of $(\varphi^0, \varphi^1, \varphi^2)$. From the reference bases of 0 and $\pi$ in a single MZI, equation (13) shows three possible bases, where $\varphi^1$ indicates the reference basis. Because the phase basis $\varphi^2$ relates with $e^{i2q}$, $\varphi^2 = (\pi/2)m$ is achieved. Thus, the phase basis set of $\varphi^2$ is represented as $\varphi^2 \in \left\{0, \frac{\pi}{2}, \pi\right\}$.

(ii) q=3 for a triply coupled MZI

$$\begin{bmatrix} 0 \\ \varphi \end{bmatrix}_1 \otimes \begin{bmatrix} 0 \\ \varphi \end{bmatrix}_2 \otimes \begin{bmatrix} 0 \\ \varphi \end{bmatrix}_3 = \begin{bmatrix} 0 \\ \varphi \\ \varphi \\ \varphi^2 \\ \varphi \\ \varphi^2 \\ \varphi^2 \\ \varphi^3 \end{bmatrix} \to (1,3,3,1). \tag{14}$$

Equation (14) shows four phase bases with two additional ones at $\pi/3$ and $2\pi/3$ as shown by two green dots in Fig. 2: $\varphi^3 = (\pi/3)m$. These added ones are equidistance located between 0 and $\pi$ bases.

(iii) q=4 (n=2)

$$\begin{bmatrix} 0 \\ \varphi \end{bmatrix}_1 \otimes \begin{bmatrix} 0 \\ \varphi \end{bmatrix}_2 \otimes \begin{bmatrix} 0 \\ \varphi \end{bmatrix}_3 \otimes \begin{bmatrix} 0 \\ \varphi \end{bmatrix}_4 \to (1,4,6,4,1). \tag{15}$$

Equation (15) is for a double ACD-MZI composed of four MZIs, resulting in three more bases added as shown by three black dots in Fig. 2. As will be analyzed further below, the basis increase according to q (MZI) is linear to with respect to q:

(iv) q=5

$$\begin{bmatrix} 0 \\ \varphi \end{bmatrix}_1 \otimes \begin{bmatrix} 0 \\ \varphi \end{bmatrix}_2 \otimes \begin{bmatrix} 0 \\ \varphi \end{bmatrix}_3 \otimes \begin{bmatrix} 0 \\ \varphi \end{bmatrix}_4 \otimes \begin{bmatrix} 0 \\ \varphi \end{bmatrix}_5 \to (1,5,10,10,5,1). \tag{16}$$

(v) q=6 (n=3)

$$\begin{bmatrix}0\\\varphi\end{bmatrix}_1 \otimes \begin{bmatrix}0\\\varphi\end{bmatrix}_2 \otimes \begin{bmatrix}0\\\varphi\end{bmatrix}_3 \otimes \begin{bmatrix}0\\\varphi\end{bmatrix}_4 \otimes \begin{bmatrix}0\\\varphi\end{bmatrix}_5 \otimes \begin{bmatrix}0\\\varphi\end{bmatrix}_6 \rightarrow (1,6,15,20,15,6,1). \quad (17)$$

(vi) q=7 (see the Supplementary Information)

$$\begin{bmatrix}0\\\varphi\end{bmatrix}_1 \otimes \begin{bmatrix}0\\\varphi\end{bmatrix}_2 \otimes \begin{bmatrix}0\\\varphi\end{bmatrix}_3 \otimes \begin{bmatrix}0\\\varphi\end{bmatrix}_4 \otimes \begin{bmatrix}0\\\varphi\end{bmatrix}_5 \otimes \begin{bmatrix}0\\\varphi\end{bmatrix}_6 \otimes \begin{bmatrix}0\\\varphi\end{bmatrix}_7 \rightarrow (1,7,21,35,35,21,7,1). \quad (18)$$

These ordered basis notations in equations (13)-(18) represent that the maximum number of the ordered phase bases is exactly the same as the number q of MZIs, where these relations are categorized in Table 1 resulting in Pascal's triangle. Here, the highest phase basis represents CBW in an asymmetric structure of MZIs of Fig. 1.

**Table 1.** The number of MZI "n" vs. the phase order "$\varphi^q$" of the output fields in equation (1).

| q \ $\varphi^q$ | **0** | **1** | **2** | **3** | **4** | **4** | **6** | **7** |
|---|---|---|---|---|---|---|---|---|
| **2 (n=1)** | 1 | 2 | 1 | | | | | |
| **3** | 1 | 3 | 3 | 1 | | | | |
| **4 (n=2)** | 1 | 4 | 6 | 4 | 1 | | | |
| **5** | 1 | 5 | 10 | 10 | 5 | 1 | | |
| **6 (n=3)** | 1 | 6 | 15 | 20 | 15 | 6 | 1 | |
| **7** | 1 | 7 | 21 | 35 | 35 | 21 | 7 | 1 |

For the q^th ordered phase basis of $\varphi^q$, the output field's intensity has a sinusoidal oscillation in the form of $e^{iq\varphi}$. Thus, the phase basis of $\varphi^q = (\pi/q)m$ is successfully summarized in Table 1. The increased phase bases, of course, result in a q-times enhanced phase resolution applicable to quantum sensing. As a result, the effective wavelength of CBW is $\lambda_{CBW} = \lambda_0/2q$, corresponding to that of PBW in $\lambda_{PBW} = \lambda_0/4N$, where N is the number of entangled photons on a BS. The 1/2 ratio between CBW and PBW is due to the $g^{(1)}$ correlation in CBW over the $g^{(2)}$ correlation in PBW. Considering the $g^{(1)}$ correlation in CBW, q functions as N. Thus, the coupling effect of a tensor matrix in equation (2) has been demonstrated for the $q\varphi$ phase oscillation, resulting in a q-times enhanced phase resolution in the output fields [17-20,28,29,35,36].

*Numerical approach*

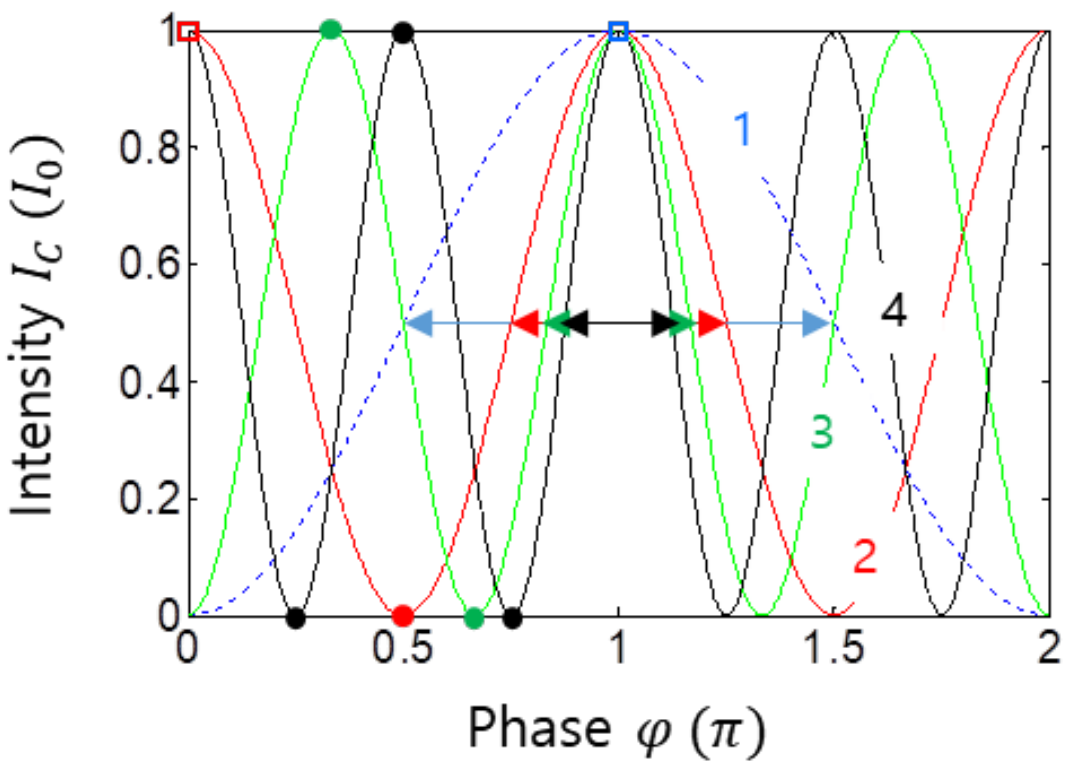

**Figure 2.** Numerical calculations for the output intensity of ACD-MZI of Fig. 1. q=1 (dotte), q=1 (red), q=3 (green), and q=4 (black). For simplicity $\varphi_B = \varphi_D = 0$ is set for $\varphi_C = -\varphi_A$ to satisfy the asymmetry coupling in CBW with $\psi = 0$ [28]. $I_C$ equivalent to $(I_A)^n$ in the n-coupled MZI in Fig. 1. The colored arrows indicate each full width at half maximum.

Figure 2 shows numerical calculations for $I_C$ for different q, where q is the basic building block of MZI satisfying ACD-MZI with $\varphi_{CD} = -\varphi_{AB}$ for $\psi = 0$ or $\varphi_{CD} = \varphi_{AB}$ for $\psi = \pi$: n=2q. The dotted curve is for a single MZI as a reference to indicate the classical diffraction limit [1]. The red curve is for q=2 showing three phase bases as shown in equation (13) (see red dot). As q increases, the number of bases increases from q=2 (red) to q=4 (black) via q=3 (green). The q-dependent phase bases in Fig. 2 show the same relation with equations (13)-(18) and Table 1. Thus, the tensor analysis in the coupled MZI system strongly supports the origin of nonclassicality in terms of newly generated phase bases in $e^{i\varphi_q}$ harmonic waves. This quantum feature of $\lambda_{CBW} = \lambda_0/2q$ is created from purely coherence optics via macroscopic coupling of the path superposition among MZIs in Fig. 1.

Instead of the serial connection of the two-mode MZIs in Fig. 1 for CBW as denoted in Table 1 and equations (13)-(18), the recursive structure of Fig. 1 can be obtained for a cavity CBW. Figure 3(a) shows a schematic of a cavity CBW, where the asymmetric phase relation between consecutive MZIs are satisfied with $\varphi_A = \varphi_D$ $(= \varphi)$ and $\varphi_B = \varphi_C$ (=0). The control phase ψ is set to zero automatically and precisely due to the shared paths located at both ends of the bow-tie structure. The alternative sign change among the ordered fields' amplitudes of $E_C$ in equation (8) results in a complete cancellation of the output fields due to superposition for all phases except for φ = π for $I_C$ if n ≫ 1, as shown in Figs. 3(b) and (c). Even with an optical loss η in the output coupler, the gradual intensity decrease in the bounced fields, however, does not affect the overall fields' cancellation (see Section 2 of the Supplementary Information). In other words, Fig. 3(b) is achieved regardless of the η value. Figure 3(a) has already been applied for CBW Sagnac interferometer in a modified scheme [37], where the highly sensitive phase resolution can be used for inertial navigation without the help of the global positioning system.

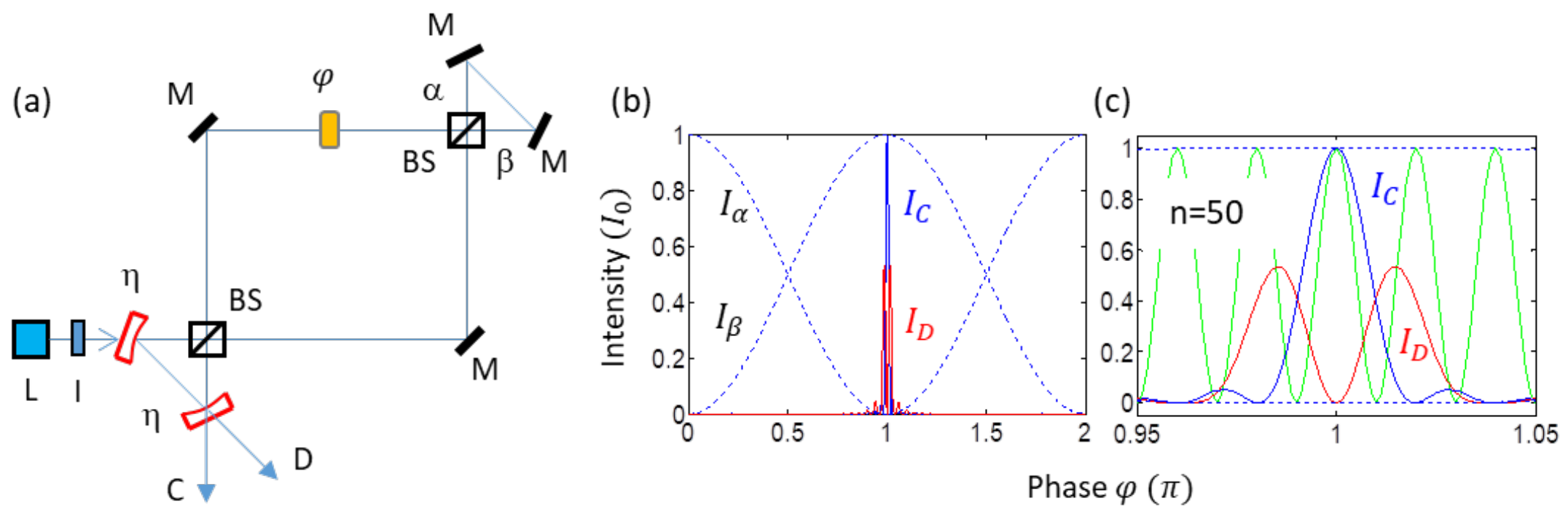

**Figure 3.** Cavity CBW. (a) Schematic of cavity CBW. (b) and (c) Numerical calculations for (a). LD, laser diode; I, isolator; η, Reflection coefficient of an optical cavity; M, mirror; BS, beam splitter; α/β: optical fields; n=number of ACD-MZI in Fig. 1. $I_\alpha/I_\beta$: output intensity of $MZI_1$. Green curve in (c): $I_C$ from Fig. 1 for n=50.

The spectral width $I_C$ in Fig. 3(c) is quite different from that of an optical cavity whose spectral width ratio ζ to the free spectral range is ζ = 0.02 with $\eta^2 = 0.98$: For low values of η~0.5, both the spectral width ratio and modulation depth of the output fields are ~50% in conventional optical cavity [1]. For the present cavity CBW in Fig. 3, however, the spectral width ratio is proportional to $\lambda_{CBW}/\lambda$, where $\lambda$ is the given wavelength, and the order of $\lambda_{CBW}$ is infinite with a clear modulation depth regardless of η (see Section 2 of the Supplementary Information). In Fig. 3(c), FWHM of $I_C$ is ~0.02π determined by $\lambda_{CBW}$ $(=\frac{\lambda}{100} for\ n = 50)$, resulting in ζ = 0.02. For n=300, $\lambda_{CBW} = \lambda/600$ and the FWHM of $I_C$ is ~0.005π, even with a $10^{-3}$ reduced amplitude for the $300^{th}$ $\lambda_{CBW}$. Because effective n should be decided by the cavity decay time in a conventional optical cavity and thus ζ = 0.005 is not possible due to low finesse with η = 0.99 [1], the role of finesse is negligible in the present cavity CBW. A high-resolution spectrometer based on Fig. 3 can also be implemented using silicon waveguides of the serial MZI structure of Fig. 2 [7,44,45].

Figure 4 shows numerical calculations for the output fields as functions of φ and ψ. Both phases have the same modulus of 2π. As mentioned above the control phase ψ functions as a toggle switch between CBW and USCKD depending on the choice of its basis. The phase sensitivity increases as n increases. The bottom row is an extension of the third row, where the width of 0.02π is confirmed in both φ and ψ. As shown in the lower right corner, the output intensity $I_D$ is quite sensitive to the phase variance of both φ and ψ. As mentioned above in Fig. 3, the number n in the cavity CBW is much bigger than 50 even with η = 0.9. Thus, the CBW sensitivity is much higher than that of optical cavity relying on finesse [1]. Most of all, the variance of ψ is zero due to the shared path in the bow-tie structure of Fig. 3. The phase noise of φ is also nearly zero in the silicon waveguide structure whose refractive index varies by temperature only [ ].

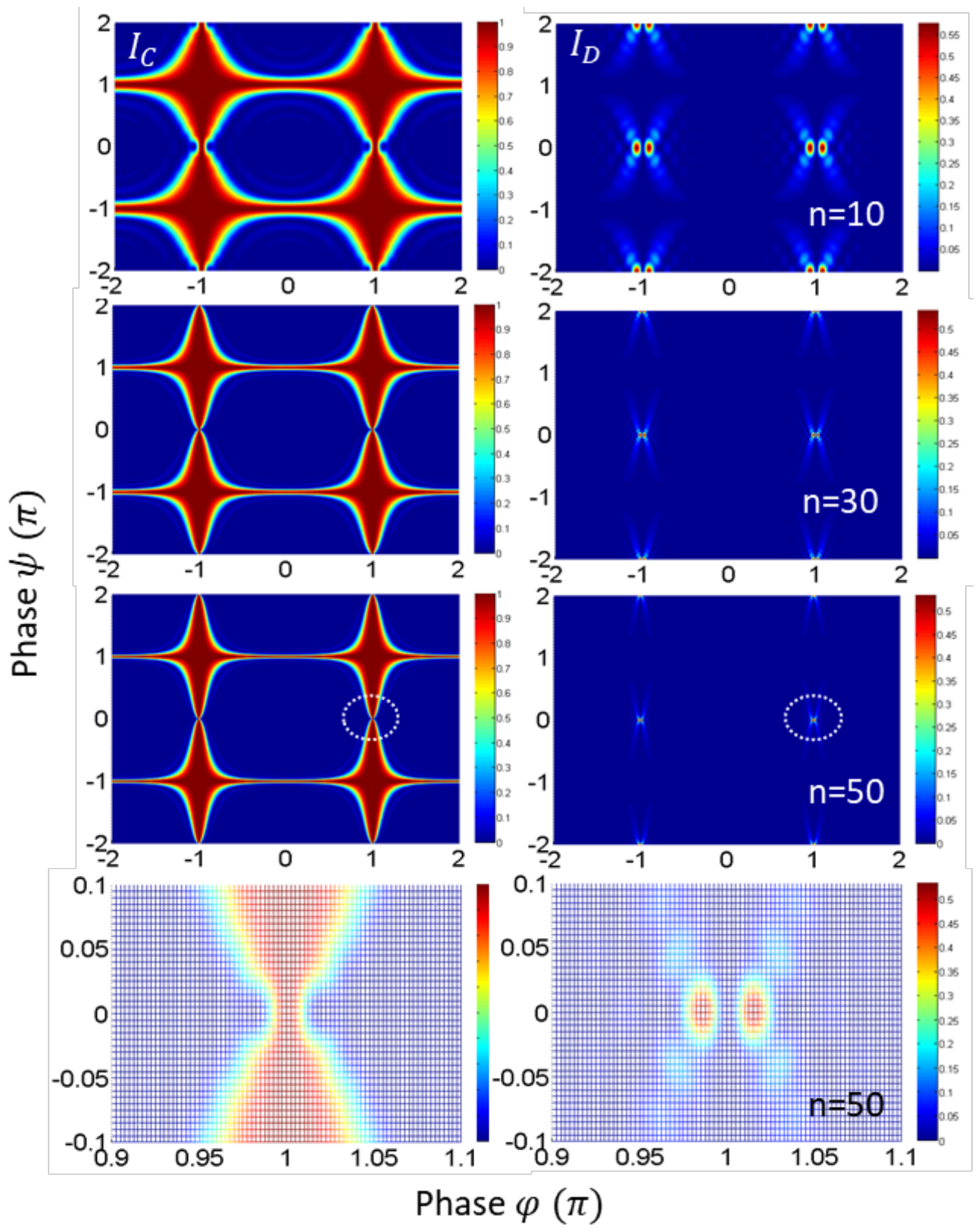

**Figure 4.** Numerical calculation of CBW in Fig. 1 as functions of φ and ψ. η = 0.9. The bottom row is the extension of the circles in the above row. The left (right) column is for $I_C$ ($I_D$).

## 3. Conclusion

In conclusion, a nonclassical feature of CBWs is analyzed using tensor matrices of binary basis systems of MZI. Analyzing a coupled MZI system using both phase basis tensor matrices and numerical calculations, the origin of the nonclassical properties of CBWs was demonstrated for the $n^{th}$ order phase basis whose phase resolution is n-times enhanced compared with the classical diffraction limit. The number of phase bases in an ACD-MZI system was also demonstrated and resulted in Pascal's triangle. For potential applications of CBWs, cavity CBWs ware also analyzed for enhanced phase resolution even with low finesse cavity. Although the output mode looks similar to the conventional optical cavity mode, its physics is completely different, satisfying nonclassical physics. For potential applications, silicon photonics may be a good tool to implement cavity CBWs with even low finesse.

## Acknowledgments

BSH acknowledges that the present research was supported by a GRI grant of GIST 2020.